\documentclass[graybox]{svmult}
\usepackage{mathptmx}       % selects Times Roman as basic font
\usepackage{helvet}         % selects Helvetica as sans-serif font
\usepackage{courier}        % selects Courier as typewriter font
\usepackage{makeidx}         % allows index generation
\usepackage{graphicx}        % standard LaTeX graphics tool
\usepackage{multicol}        % used for the two-column index
\usepackage[bottom]{footmisc}% places footnotes at page bottom
\usepackage{amsmath,amssymb,amsfonts}        % AMS math packages
\makeindex             % used for the subject index
\usepackage{amsmath, latexsym, amsfonts,bm,amssymb,mathscinet} % Useful basic packages.
\usepackage{mathtools}
\usepackage{xcolor}

\PassOptionsToPackage{hyphens}{url} %inserted on 3 June 2022
\usepackage[plainpages=false,breaklinks]{hyperref} %inserted on 3 June 2022

\usepackage{enumerate}
\usepackage{units} % nicefrac
\usepackage[ruled,vlined]{algorithm2e}
\newcommand{\ind}{\mathbf{1}}

\renewcommand{\th}{\theta}

\newcommand{\EE}{\mathbb{E}}
\newcommand{\VV}{\mathbb{V}}
\newcommand{\var}{\mathbb{V}\mathrm{ar}}

\def\epsilon{\varepsilon}

\def\dd{{ {\mathrm{d}}}}
\renewcommand{\th}{\theta}
\newcommand{\ig}{\operatorname{IG}}

\allowdisplaybreaks

\usepackage{color}

\definecolor{DarkGoldenrod}{rgb}{0.78, 0.45, 0.04}

\graphicspath{{./figs_pdf/}{./}}

\begin{document}

\title*{Nonparametric {B}ayesian volatility learning under microstructure noise}
\author{Shota Gugushvili, Frank van der Meulen, Moritz Schauer, and Peter Spreij}
\authorrunning{Gugushvili et al.}
\institute{Shota Gugushvili 
\at Biometris, Wageningen University \& Research, 
Postbus 16, 6700 AA Wageningen, 
The Netherlands. \email{shota@yesdatasolutions.com}
\and Frank van der Meulen \at Department of Mathematics, Vrije Universiteit Amsterdam,
De Boelelaan 1111,
1081 HV Amsterdam,
The Netherlands. \email{f.h.van.der.meulen@vu.nl}
\and Moritz Schauer \at 
Department of Mathematical Sciences, Chalmers University of Technology and University of Gothenburg, SE-412 96 G\"{o}teborg, Sweden. \email{smoritz@chalmers.se}
\and Peter Spreij \at Korteweg-de Vries Institute for Mathematics,
University of Amsterdam,
P.O. Box 94248,
1090 GE Amsterdam,
The Netherlands and Institute for Mathematics, Astrophysics and Particle Physics, Radboud University, Nijmegen, The Netherlands.~\email{spreij@uva.nl}} 
\maketitle

%%Orcid:
%https://orcid.org/0000-0002-6963-295X %shota
%https://orcid.org/0000-0001-7246-8612 %frank
%https://orcid.org/0000-0003-3310-7915 %moritz
%https://orcid.org/0000-0002-6416-6320 %Peter

\abstract{In this work, we study the problem of learning the volatility under market microstructure noise. Specifically, we consider noisy discrete time observations from a  stochastic differential equation and develop a novel computational method to learn the diffusion coefficient of the equation.  We take a nonparametric Bayesian approach, where we \emph{a priori} model the volatility function as piecewise constant. Its prior is specified via the inverse Gamma Markov chain. Sampling from the posterior  is accomplished by incorporating the Forward Filtering Backward Simulation algorithm in the Gibbs sampler. Good performance of the method is demonstrated on two representative synthetic data examples. We also apply the method on a EUR/USD exchange rate dataset. Finally we present a limit result on the prior distribution.
}


\section{Introduction}
\label{section:problem}

%\subsection{Problem formulation}
Let the one-dimensional stochastic differential equation (SDE)
\begin{equation}
\label{sde}
\dd X_t=b(t,X_t)\,\dd t+s(t)\,\dd W_t, \quad X_0=x_0, \quad t\in[0,T],
\end{equation}
be given. Here $W$ is a standard Wiener process and $b$ and $s$ are referred to as the  drift function and volatility function, respectively.
We assume (not necessarily uniformly spaced) observation times $\{t_i: 1\le i\le n\}$ and observations
$
\mathcal{Y}_n =\{Y_1, \ldots, Y_{n}\},
$
where
\begin{equation}
\label{eq:obs}
Y_{i}=X_{t_{i}}+V_{i}, \quad 0<t_1<\cdots<t_n=T,
\end{equation}
and $\{ V_i \}$ is a sequence of independent and identically distributed random variables, independent of $W$.
Our aim is to learn  the volatility $s$ using the noisy observations $\scr{Y}_n$. Knowledge of the volatility is of paramount importance in financial applications, specifically in pricing financial derivatives, see, e.g., \cite{musiela05}, and in risk management. 

The quantity $\Delta t_i = t_i - t_{i-1}$ is referred to as the observation density, especially if the time instants are equidistant, and then $1/\Delta t_i$ will be the frequency of the observations. 
Small values of $\Delta t_i$ correspond to high frequency, dense-in-time data. Intraday financial data are commonly thought to be high frequency data. In this high frequency financial data setting, which is the one we are interested in the present work, the measurement errors $\{V_{i}\}$ are referred to as  microstructure noise. Their inclusion in the model aims to reflect such features of observed financial time series as their discreteness or approximations due to market friction. Whereas for low-frequency financial data these can typically be neglected without much ensuing harm, empirical evidence shows that this is not  the case for high frequency data; cf.~\cite{mykland12}.

There exists a large body of statistics and econometrics literature on nonparametric volatility estimation under  microstructure noise. See, e.g.,
\cite{hoffmann12},
\cite{jacod09}, \cite{mykland:zhang:09},
\cite{reiss11}, \cite{sabel15}; a recent overview is \cite{mykland12}. The literature predominantly deals with estimation of the integrated volatility $\int_0^t s^2(u)\dd u$, although inference on $s$ has also been studied in several of these references. Various methods proposed in the above-mentioned works share the property of being \emph{frequentist} in nature. An important paper \cite{mancini2015spot}  is a mix of theoretical and practical results, with a clear predominance of the former. Its main purpose (see page 262) is proposing a unifying frequentist approach to spot volatility estimation, from which many previously existing approaches would be derived as special cases. This allows comparison between them.  As our approach is Bayesian, our results do not fall under the general umbrella of \cite{mancini2015spot}.

In this paper, our main and novel contribution is the development of a \emph{practical} nonparametric Bayesian approach to volatility learning under microstructure noise.
We specify an inverse Gamma Markov chain prior (Cf.\ \cite{cemgil07}) on the volatility function $s$ and reduce our model to the Gaussian linear state space model. Posterior inference in the latter is performed using Gibbs sampling  including  a Forward Filtering Backward Simulation (FFBS) step.
We demonstrate good performance of our method on two representative simulation examples. The first example uses a benchmark function, popular in nonparametric regression, see~\cite{fan95}, as the volatility function. In the second example we consider a well known and widely used stochastic volatility model, the Heston model, see~\cite{heston93}, or \cite[Section~10.3.3]{filipovic2009} and \cite[Chapter~19, Appendix~A]{brigomercurio2006}. In both examples our approach shows  accurate results.
We also apply our method to a real data set of  EUR/USD exchange rates and we deduce a clear and understandable variation in the volatility over time.

In general, a nonparametric approach reduces the risk of model misspecification; the latter may lead to distorted inferential conclusions. The presented nonparametric method is not only useful for an exploratory analysis of the problem at hand (cf.~\cite{silverman86}), but also allows honest representation of inferential uncertainties (cf.~\cite{mueller13}). Attractive features of the Bayesian approach include its internal coherence, automatic uncertainty quantification in parameter estimates via Bayesian credible sets, and the fact that it is a fundamentally likelihood-based method.
For a modern monographic treatment of nonparametric Bayesian statistics see \cite{ghosal17}; an applied perspective is found in \cite{mueller15}.

The paper is organised as follows: 
in Section~\ref{section:approach} we introduce in detail our approach, followed by Section~\ref{addresults}  where the limiting behaviour of the prior on the squared volatility is derived under mesh refinement. 
In Section~\ref{section:simulations} we test its practical performance on synthetic data examples. Section~\ref{section:real} applies our method on a real data example. Section~\ref{section:conclusions} summarises our findings. Finally, Appendix~\ref{appendixB} gives further implementational details.

\subsection*{Notation}

We denote the inverse Gamma distribution with shape parameter $\alpha>0$ and scale parameter $\beta>0$ by  $\ig(\alpha,\beta)$. Its density is
\[
x \mapsto \frac{\beta^{\alpha}}{\Gamma(\alpha)} x^{-\alpha-1} e^{-\beta/x}, \quad x>0.
\]
By $N(\mu,\sigma^2)$ we denote a normal distribution with mean $\mu\in\mathbb{R}$ and variance $\sigma^2>0$. The uniform distribution on an interval $[a,b]$ is denoted by $\operatorname{Uniform}(a,b).$ For a random variate $X$, the notation $X\sim p$ stands for the fact that $X$ is distributed according to a density $p$, or is drawn according to a density $p$. Conditioning of a random variate $X$ on a random variate $Y$ is denoted by $X\mid Y$. By $\lfloor x \rfloor$ we denote the integer part of a real number $x$. The notation $p \propto q$ for a density $p$ denotes the fact that a positive function $q$ is an unnormalised density corresponding to $p$: $p$ can be recovered from $q$ as $q/\int q$. Finally, we use the shorthand notation $a_{k:\ell}=(a_k,\ldots,a_{\ell})$.

%%%%%%%%%%%%%%%%%%%%%%%%%%%%%%%%%%%%%%%%%%

\section{Methodology}
\label{section:approach}

In this section we introduce our  methodology for inferring the  volatility. We first recast the model into a simpler form that is amenable to computational analysis, next specify a nonparametric prior on the volatility, and finally describe an MCMC method  for sampling from the posterior. 

\subsection{Linear state space model}

Let $t_0=0$. By equation \eqref{sde}, we have
\begin{equation}
\label{eq:sde}
X_{t_{i}}=X_{t_{i-1}}+\int_{t_{i-1}}^{t_{i}}b(t,X_t)\dd t+\int_{t_{i-1}}^{t_{i}} s(t)\dd W_t.
\end{equation}
We derive our method under the assumption that  the ``true'', data-generating volatility $s$ is a deterministic function of time $t$.  Next, if the ``true'' $s$ is in fact a stochastic process, we apply our procedure without further changes, as if $s$ were deterministic. As shown in the example of Subsection~\ref{subsec:heston}, this works in practice. 
 That this is the case is easiest to understand in the situation where one can discern a two-stage procedure. First the stochastic volatility is generated, and given a realization of it, the observations are generated by an independent Brownian motion $W$. In 
\cite{kanayakristensen2016} such an approach is used for simulation.

Over short time intervals $[{t_{i-1}},{t_{i}}]$, the term $\int_{t_{i-1}}^{t_{i}} s(t)\dd W_t$ in \eqref{eq:sde}, roughly speaking, will dominate the term  $\int_{t_{i-1}}^{t_{i}}b(t,X_t)\dd t$, as the latter scales as $\Delta t_i$, whereas the former as $\sqrt{\Delta t_i}$ (due to the properties of the Wiener process paths). As our emphasis is on learning $s$ rather than $b$, following \cite{gugu2020}, \cite{gugu18} we act as if the process $X$ had a zero drift, $b\equiv 0$. The justification of this procedure is explained \cite{gugu2020}.
A similar idea is often used in \emph{frequentist} volatility estimation procedures in the {high frequency} financial data setting; see \cite{mykland12}, Section 2.1.5 for an intuitive exposition. Formal results why this works in specific settings rely on Girsanov's theorem, see, e.g., \cite{gugu2020}, \cite{hoffmann12}, \cite{mykland:zhang:09}. Further reasons why one would like to set $b=0$ are that   $b$ is a \emph{nuisance parameter}, in specific applications its appropriate parametric form might be unknown, and finally, a single observed time series is not sufficient to learn $b$ consistently (see~\cite{ignatieva12}).

We thus \emph{assume}
$
X_{t_{i}}=X_{t_{i-1}}+U_{i},
$
where 
$
U_{i}=\int_{t_{i-1}}^{t_{i}} s(t)\dd W_t.
$
Note that then
\begin{equation}
\label{ui}
U_{i} \sim N(0,w_i)\quad \text{with} \quad w_i=\int_{t_{i-1}}^{t_{i}} s^2(t)\dd t,
\end{equation}
and also that $\{U_{i}\}$ is a  sequence of independent random variables. To simplify our notation, write $x_i=X_{t_{i}}$, $y_i=Y_{i}$, $u_i=U_{i}$, $v_i=V_{i}$. The preceding arguments and \eqref{eq:obs} allow us to reduce our model to the \emph{linear state space model}
\begin{equation}
\label{eq:ssm}
\begin{split}
x_i&=x_{i-1}+u_i,\\
y_i&=x_i+v_i,
\end{split}
\end{equation}
where $i = 1, \dots, n$.
The first equation in \eqref{eq:ssm} is the \emph{state equation}, while the second equation is the \emph{observation equation}. We assume that $\{v_i\}$ is a sequence of independent $N(0,\eta_v)$ distributed random variables,  independent of the Wiener process $W$ in \eqref{sde}, so that $\{v_i\}$ is independent of $\{u_i\}$.  For justification of such assumptions on the noise sequence $\{v_i\}$ from a practical point of view, see \cite{sabel15}, page 229. We endow the \emph{initial state} $x_0$ with the $N(\mu_0,C_0)$ prior distribution. Then \eqref{eq:ssm} is a \emph{Gaussian} linear state space model. This is  very convenient computationally. Had we not followed this route, we would have had to deal with an \emph{intractable likelihood}, which constitutes the main \emph{computational bottleneck} for Bayesian inference in SDE models; see, e.g,  \cite{papa13} and \cite{vdm-s-estpaper} for discussion.

\subsection{Prior}
\label{sec:prior}

For the measurement error variance $\eta_v$, we assume a priori $\eta_v \sim \ig(\alpha_v,\beta_v)$. The construction of the prior for $s$ is more complex and follows \cite{gugu18}, that in turn relies on \cite{cemgil07}. Fix an integer $m<n$. Then we have a unique decomposition $n=mN+r$  with $0\leq r<m$, where $N=\lfloor {n}/{m}\rfloor$. Now define bins $B_k=[t_{m(k-1)},t_{mk})$, $k=1,\ldots,N-1$, and $B_N=[t_{m(N-1)},T]$.
We model $s$ as
\begin{equation}\label{eq:prior-s}
s=\sum_{k=1}^{N} \xi_k \ind_{B_k}, 
\end{equation}
where  $N$ (the number of bins) is a hyperparameter. Then $
s^2=\sum_{k=1}^{N} \theta_k \ind_{B_k},
$
where  $\theta_k=\xi_k^2$. We complete the prior specification for  $s$ by  assigning a prior distribution to the coefficients $\theta_{1:N}$. For this purpose, we introduce auxiliary variables $\zeta_{2:N}$, and suppose the sequence $\theta_1,\zeta_2,\theta_2,\ldots,\zeta_k,\theta_k,\ldots,\zeta_N,\theta_N$ forms a Markov chain (in this order of variables). The transition distributions of the chain are defined  by
\begin{equation}
\label{formula:prior}
\theta_1 \sim \ig(\alpha_1,\beta_1), \quad \zeta_{k+1} | \theta_k \sim \ig(\alpha,\alpha \theta_k^{-1}),  \quad \theta_{k+1} | \zeta_{k+1} \sim \ig(\alpha,\alpha \zeta_{k+1}^{-1}),
\end{equation}
where $\alpha_1,\beta_1,\alpha$ are hyperparameters. We refer to this chain as an inverse Gamma Markov chain, see \cite{cemgil07}. The corresponding prior on $\theta_{1:N}$ will be called the inverse Gamma Markov chain (IGMC) prior. The definition in \eqref{formula:prior} ensures that $\theta_1,\ldots,\theta_N$ are positively correlated, which imposes \emph{smoothing} across different bins $B_k$. Simultaneously, it ensures partial conjugacy in the Gibbs sampler that we derive below, leading to simple and tractable MCMC inference.
In our experience, an uninformative choice $\alpha_1,\beta_1\rightarrow 0$ performs well in practice. We also endow $\alpha$ with a prior distribution and assume $\log \alpha\sim N(a,b)$, with hyperparameters $a\in\mathbb{R},b>0$ chosen so as to render the hyperprior on $\alpha$ diffuse. As explained in \cite{gugu2020}, \cite{gugu18}, the hyperparameter $N$ (or equivalently $m$) can be considered both as a smoothing parameter and the resolution at which one wants to learn the volatility function. Obviously, given the limited amount of data, this resolution cannot be made arbitrarily fine. On the other hand, as shown in \cite{gugu18} (see also \cite{gugu:ppp:18}), inference with the IGMC prior is quite robust with respect to a wide range of values of $N$, as the corresponding Bayesian procedure has an additional \emph{regularisation parameter} $\alpha$ that is learned from the data. Statistical optimality results in \cite{munk:schmidt10} suggest that in our setting $N$  should be chosen considerably smaller than in the case of an SDE observed without noise (that was studied via the IGMC prior in \cite{gugu18}).

\subsection{Likelihood}
\label{section:likelihood}
Although an expression for the posterior of $s$ can be written down in closed form, it is not amenable to computations. This problem is alleviated by following a data augmentation approach, in which  $x_{0:n}$ are treated as  \emph{missing data}, whereas the $y_{1:n}$ are the observed data; cf.~\cite{tanner87}. An expression for the joint density of all random quantities involved is easily derived from the prior  specification and \eqref{eq:ssm}. We have
\begin{align*}
 \lefteqn{p(y_{1:n}, x_{0:n}, \th_{1:N}, \zeta_{2:N}, \alpha, \eta_v)  =} \\
&\qquad \left(\prod_{k=1}^n p(y_k\mid x_k, \eta_v)\right)  p(x_{0:n} \mid \th_{1:N}) \\
&\qquad \times p(\th_1)  \prod_{k=1}^{N-1} \left[ p(\zeta_{k+1}\mid \th_k, \alpha) p(\th_{k+1} \mid \zeta_{k+1}, \alpha)\right] p(\alpha) p(\eta_v).
\end{align*}
Except for $p(x_{0:n} \mid \th_{1:N})$, all the densities here have been specified in the previous subsections. To obtain an expression for the latter, define (with $\Delta_i\equiv \Delta t_i$)
\begin{align*}
Z_k &= \sum_{i=(k-1)m+1}^{km} \frac{(x_i-x_{i-1})^2}{\Delta_i}, \quad k=1,\ldots,N-1,\\
Z_N &= \sum_{i=(N-1)m+1}^{n} \frac{(x_i-x_{i-1})^2}{\Delta_i},
\end{align*}
and set $m_k=m$ for  $k=1,\ldots,N-1$, and $m_N=m+r$. Then 
\[ p(x_{0:N} \mid  \th_{1:N}) \propto e^{(x_0-\mu_0)^2/(2C_0)} \prod_{k=1}^N  \th_k^{-m_k/2} \exp\left(-\frac{ Z_k}{2\th_k} \right). \]

\subsection{Gibbs sampler}

We use the Gibbs sampler to sample from the joint conditional distribution of $(x_{0:n}, \th_{1:N}, \zeta_{2:N}, \eta_v, \alpha)$ given $y_{1:n}$. The full conditionals of $\th_{1:N}$, $\zeta_{2:N}$, $\eta_v$  are easily derived from Section~\ref{section:likelihood} and recognised to be of the inverse Gamma type, see Subsection~\ref{section:drawing}. The parameter $\alpha$ can be updated via a Metropolis-Hastings step. For updating $x_{0:N}$, conditional on all other parameters, we  use the standard Forward Filtering Backward Simulation (FFBS) algorithm for Gaussian state space models (cf.\ Section 4.4.3 in \cite{petris09}). The  resulting Gibbs sampler is summarised in Algorithm~\ref{pseudocode}. For details, see Appendix~\ref{appendixB}.

\begin{algorithm}%[H]
\SetAlgoLined
\KwData{Observations $y_{1:n}$}
Hyperparameters $\alpha_1,$ $\beta_1$, $\alpha_v$, $\beta_v$, $a$, $b$, $N$\;
\KwResult{Posterior samples $\theta_{1:N}^{i}:i=1,\ldots,M$}
 Initialization $\theta_{1:N}^0,$ $\zeta_{1:N}^0$, $\eta_v^0$, $\alpha^0$\;
\While{$i \leq M$}{
    sample $x_{0:n}^i$ via FFBS\;
sample $\theta_{1:N}^i$ from the inverse Gamma full conditionals\;
sample $\zeta_{2:N}^i$ from the inverse Gamma full conditionals\;
sample $\eta_v^i$ from the inverse Gamma full conditional\;
sample $\alpha^i$ via a Metropolis-Hastings step\;
set $i=i+1$. 
    }
 \caption{Gibbs sampler for volatility learning}
\label{pseudocode}
\end{algorithm}

\section{Asymptotics for the prior on the squared volatility}
\label{addresults}

We first provide in Proposition~\ref{prop:tmv} results on the prior conditional mean and variance of the $\theta_k$ as in Section~\ref{sec:prior}. These results will be exploited to find an asymptotic regime for the $\theta_k$ when the number of bins tends to infinity. In this section we depart from the original setting with $\alpha$ as a random hyperparameter  (having a lognormal  distribution), but instead we take it as a deterministic one that we let grow to infinity to obtain our asymptotic results.

\begin{proposition}\label{prop:tmv}
If $\alpha>2$ is a fixed parameter then 
the IGMC prior of Section~\ref{sec:prior} satisfies
\begin{align}
\EE_k [\theta_{k+1}-\theta_k] 
& =  \frac{1}{\alpha-1}\theta_k, \label{eq:cmean} \\
\VV_k (\theta_{k+1}-\theta_k) 
& =  \frac{\alpha(2\alpha-1)}{(\alpha-1)^2(\alpha-2)} \theta_k^2, \label{eq:cvar}
\end{align}	
where $\EE_k$ and $\VV_k$ respectively denote expectation and  variance, conditional on $\theta_k$. Consequently, the conditional mean squared error $\EE_k(\theta_{k+1}-\theta_k)^2$ equals $\frac{2(\alpha+1)\theta_k^2}{(\alpha-1)(\alpha-2)}$.
\end{proposition}

\begin{proof}
We will use that for $Z\sim \ig(a,b)$ it holds that
\begin{align}\label{eq:mean_iginv}
\EE Z^{-1} & = \frac{a}{b}, \\
\EE Z^{-2} & = \frac{a(a+1)}{b^2}.\nonumber
\end{align}
First we consider the conditional mean. 
Since $\theta_{k+1} \mid \zeta_{k+1} \sim \ig(\alpha, \alpha \zeta_{k+1}^{-1})$ we have
\begin{equation}\label{eq:condmean}
 \EE[\theta_{k+1} \mid \zeta_{k+1}] =\frac{\alpha \zeta_{k+1}^{-1}}{\alpha-1} 
 \end{equation}
provided $\alpha >1$. Exploiting that the sequence $\theta_1,\zeta_2,\theta_2,\ldots,\zeta_k,\theta_k,\ldots,\zeta_N,\theta_N$ forms a Markov chain (in this order of variables), one has 
\[  
\EE_k \theta_{k+1}= \EE_k \EE [\th_{k+1} \mid \theta_k,\zeta_{k+1}]= \EE_k \EE [\th_{k+1} \mid \zeta_{k+1}]  = \frac{\alpha}{\alpha-1} \EE_k  \zeta_{k+1}^{-1},
\]
where we used \eqref{eq:condmean} at the last equality sign. Using \eqref{eq:mean_iginv} and $\zeta_{k+1}\mid \theta_k \sim \ig(\alpha, \alpha\theta_k^{-1})$ one obtains
\[ \EE_k \theta_{k+1} = \frac{\alpha}{\alpha-1} \frac{\alpha}{\alpha\theta_k^{-1}}= \frac{\alpha}{\alpha-1}\theta_k, 
\]
which is equivalent to \eqref{eq:cmean}.

Next we calculate the conditional variance. We have 
\begin{equation}\label{eq:condvar} \VV[\theta_{k+1} \mid \zeta_{k+1}] = \frac{\alpha^2 \zeta_{k+1}^{-2}}{(\alpha-1)^2(\alpha-2)} \end{equation}
 provided $\alpha>2$. 
We need the following variation on the law of total variance. If $X\in L^2(\Omega,\mathcal{F},\mathbb{P})$, and $\mathcal{G}$, $\mathcal{H}$ are subsigma-algebras of $\mathcal{F}$ with $\mathcal{H}\subset\mathcal{G}$. 
Then, with $X^\mathcal{G}=\EE[X\mid\mathcal{G}]$, it holds that
\[
\var(X\mid\mathcal{H})=\EE[\var(X\mid\mathcal{G})\mid\mathcal{H}]+\var(X^\mathcal{G}\mid\mathcal{H}).
\]
We use this result with $X=\theta_{k+1}$, $\mathcal{G}=\sigma(\theta_k,\zeta_{k+1})$, $\mathcal{H}=\sigma(\theta_k)$, obtaining
\begin{align*}
\var(\theta_{k+1}\mid\theta_k) 
& =\EE[\var(\theta_{k+1}\mid\theta_k,\zeta_{k+1})\mid\theta_k]+\var(\EE[\theta_{k+1}\mid\theta_k,\zeta_{k+1}]\mid\theta_k) \\
& =\EE[\var(\theta_{k+1}\mid\zeta_{k+1})\mid\theta_k]+\var(\EE[\theta_{k+1}\mid\zeta_{k+1}]\mid\theta_k),
\end{align*} 
as now $X^\mathcal{G}=\EE[\theta_{k+1}\mid\theta_k,\zeta_{k+1}]=\EE[\theta_{k+1}\mid\zeta_{k+1}]$ and $\var(\theta_{k+1}\mid\theta_k,\zeta_{k+1})=\var(\theta_{k+1}\mid\zeta_{k+1})$ in view of the Markov property.
Hence, in our abbreviated notation,
\[ \VV_k \theta_{k+1} = \EE_k \VV(\theta_{k+1} \mid \zeta_{k+1}) + \VV_k \EE [\theta_{k+1} \mid \zeta_{k+1}]. \]
Hence, using \eqref{eq:condmean} and \eqref{eq:condvar}
\begin{align*} \VV_k \theta_{k+1} &= \frac{\alpha^2}{(\alpha-1)^2(\alpha-2)} \EE_k \zeta_{k+1}^{-2} + \VV_k \frac{\alpha \zeta_{k+1}^{-1}}{\alpha-1} \\ & = \left( \frac{\alpha^2}{(\alpha-1)^2(\alpha-2)} + \frac{\alpha^2}{(\alpha-1)^2}\right) \EE_k \zeta_{k+1}^{-2} - \frac{\alpha^2}{(\alpha-1)^2} \left(\EE_k \zeta_{k+1}^{-1}\right)^2\\ & = \frac{\alpha^2}{(\alpha-1)(\alpha-2)} \EE_k \zeta_{k+1}^{-2} - \frac{\alpha^2}{(\alpha-1)^2} \left(\EE_k \zeta_{k+1}^{-1}\right)^2 \\ & =\frac{\alpha^2}{(\alpha-1)(\alpha-2)} \frac{\alpha(\alpha+1)}{\alpha^2 \th_k^{-2}} - \frac{\alpha^2}{(\alpha-1)^2} \left(\frac{\alpha}{\alpha\th_k^{-1}}\right)^2\\ & = \frac{\alpha(2\alpha-1)}{(\alpha-1)^2(\alpha-2)} \theta_k^2, 
\end{align*}
which establishes \eqref{eq:cvar}.	 The expression for the mean squared error follows from \eqref{eq:cmean} and \eqref{eq:cvar}.
\end{proof}

%%%%%%%%%%%%%%%%%%%%%%%%

%\subsection{Asymptotics for the prior on the squared volatility}

The expression for the conditional mean squared error in Proposition~\ref{prop:tmv} shows that it decreases in $\alpha$, which illustrates the regularising property of this parameter.
Therefore we are interested in the  behaviour of the prior distribution on the $\theta_k$ for large values of $\alpha$. In fact we will scale $\alpha$ with the number of bins $N$ to obtain a limit result by applying Donsker's theorem (Cf.\ Corollary VII.3.11 in \cite{jacod2013limit}), upon letting $N\to \infty$.  We give some heuristics to derive the limit behaviour. 

Take $\alpha= \gamma N$, where in the latter expression $\gamma$ is a positive scaling factor.
As the law of the $\theta_k$ depends on $\gamma N$, we are thus interested in the law of $\theta_k$ for $N\to\infty$. This entails simultaneously increasing the number of bins as also the dependence of the values on the bins. 
Below we argue that  under the IGMC-prior, with $\theta_1$ fixed, the process $
t\mapsto s^2(t)$, with $s^2(t)=\sum_{k=1}^{N} \theta_k \ind_{B_k}(t)$, converges weakly to the continuous time process $t\mapsto \theta_1 Z_t$ where 
\begin{equation}\label{eq:limitZ} 
\log Z_t=\sqrt{\frac{2}{\gamma}} W_t-\frac{1}{\gamma} t.
\end{equation}
The expressions for conditional mean and variance of Proposition~\ref{prop:tmv} are for large $N$ (in particular then $\alpha=\gamma N>2$) approximately equal to
\begin{align}
\EE_k \theta_{k+1}
& =  \theta_k, \label{eq:newcmean}\\
\VV_k (\theta_{k+1}) 
& =  \frac{2\theta_k^2}{\gamma N}. \label{eq:newcvar}
\end{align}
These properties are shared with $\theta_k$ generated by the recursion
\[
\theta_{k+1}=\theta_k\left(1+\sqrt{\frac{2}{\gamma N}}z_k\right),
\]
where, for all $k$, $z_k$ is independent of $\theta_1,\ldots,\theta_k$, $\EE z_k=0$, $\var\, z_k=1$. The $z_k$ can be seen as martingale differences, or even as IID random variables.
Solving the recursion starting from $\theta_1$ after taking logarithms leads to
\begin{align*}
\log \theta_{k+1}- \log \theta_1
& =\sum_{j=1}^{k}\log(1+\sqrt{\frac{2}{\gamma N}}z_j) \\
&\approx \sum_{j=1}^{k}\sqrt{\frac{2}{\gamma N}}z_j-\sum_{j=1}^{k}\frac{1}{\gamma N}z_j^2 \\
&=\sqrt{\frac{2}{\gamma}}\sqrt{\frac{1}{N}}\sum_{j=1}^{k}z_j-\frac{1}{\gamma}\frac{1}{N}\sum_{j=1}^{k}z_j^2, 
\end{align*}
where the approximation is based on a second order Taylor expansion of the logarithm and makes sense for large values of $N$.
Introduce the processes $W^N=\{W^N_t, t\in [0,1]\}$ and $A^N=\{A^N_t, t\in [0,1]\}$ by
\begin{align*}
W^N_t & =\sqrt{\frac{1}{N}}\sum_{j\leq Nt} z_j, \\
A^N_t & =\frac{1}{N}\sum_{j\leq Nt} z_j^2.
\end{align*} 
By Donsker's theorem, one has weak convergence of $W^n$ to $W$, a standard Brownian motion. 
Furthermore, $A^N$ converges uniformly in probability to $A$, $A_t=t$. It follows that $\sqrt{\frac{2}{\gamma}}W^N-\frac{1}{\gamma}A^N$ weakly converges to $\sqrt{\frac{2}{\gamma}}W-\frac{1}{\gamma}A$. Note that $A$ is the quadratic variation process $\langle W\rangle$. Hence $\sqrt{\frac{2}{\gamma}}W^N-\frac{1}{\gamma}A^N$ converges weakly to $\sqrt{\frac{2}{\gamma}}W-\frac{1}{2}\langle \sqrt{\frac{2}{\gamma}}W\rangle$.

Consequently, assuming $tN$ is an integer, for $k=tN$ one finds that the distribution of $\log\theta_k-\log \theta_0$ is approximately (for large $N$) that of $\sqrt{\frac{2}{\gamma}}W_t-\frac{1}{\gamma}t$, which is normal $N(-\frac{t}{\gamma},\frac{2t}{\gamma})=N(-\frac{k}{\gamma N},\frac{2k}{\gamma N})$. The $\theta_k$ can also approximately be generated by the recursion, yielding log-normal random variables,
 \[
\theta_{k+1}=\theta_k\exp(\xi_k),
\]
where the $\xi_k$ are IID random variables with common $N(-\frac{1}{\gamma N},\frac{2}{\gamma N})$ distribution.  Indeed for this recursion one finds $\EE_k \theta_{k+1} =  \theta_k$ and
$\VV_k (\theta_{k+1}) = \theta_k^2 (\exp(\frac{2}{\gamma N})-1)\approx\frac{2\theta_k^2}{\gamma N}$, which coincide with the earlier found expressions \eqref{eq:newcmean} and \eqref{eq:newcvar} for conditional mean and variance.

Moreover the continuous time approximation $Z$ of the $\theta_k/\theta_1$, with $Z$ as in \eqref{eq:limitZ}, is the Dol\'eans exponential $\mathcal{E}(\sqrt{\frac{2}{\gamma}} W)$ and thus satisfies the stochastic differential equation 
\begin{equation}\label{eq:sdez}
\dd Z_t=\sqrt{\frac{2}{\gamma}}Z_t\,\dd W_t.
\end{equation}
In Section~\ref{subsec:heston} we will provide results for the Heston model with the product of $\theta_1$ and this limit process as the squared volatility process.

\section{Synthetic data examples}
\label{section:simulations}

In this section we test the practical performance of our method on challenging synthetic data examples. The goal is to illustrate the ability of our method to recover the volatility in a controlled setting where the ground truth is known and thus the quality of inferential results can be assessed directly. We also show good practical performance of the method in a situation which formally does not fall under the derivations made in Section~\ref{section:approach}; see Subsection~\ref{subsec:heston} below.

The simulation setup is as follows: we take the time horizon $T=1$ and generate $n=4\, 000$ observations as follows. First, using a fine grid of $10n+1$ time points which are sampled from the $\operatorname{Uniform}(0,1)$ distribution, conditional on including 0 and 1, a realisation of the process $X$ is obtained via Euler's scheme, see~\cite{glasserman2004} or \cite{kloedenplaten1992}. The $n$ time points $\{ t_i \}$ are then taken as a random subsample of those times, conditional on including $1$. The settings used for the Gibbs sampler are given in Appendix~\ref{appendixC}. In each example below, we plot the posterior mean and $95\%$ marginal credible band (obtained from the central posterior intervals for the coefficients $\xi_k=\sqrt{\theta}_k$ in \eqref{eq:prior-s}). The former gives an estimate of the volatility, while the latter provides a means for uncertainty quantification.

All the computations are performed in the programming language Julia, see \cite{bezanson17}, and we provide the code used in our examples. The hardware and software specifications for the MacBook used to perform simulations are: CPU: Intel(R) Core(TM) M-5Y31 CPU @ 0.90GHz; OS: macOS (x86\_64-apple-darwin14.5.0).

\subsection{Fan \& Gijbels function}\label{subsec:fan}

Suppose the volatility function is given by
\begin{equation}\label{s3}
s(t)= 3/2 + \sin(2(4t-2)) + 2\exp(-16(4t-2)^2), \quad t\in[0,1].
\end{equation}
This is a popular benchmark in nonparametric regression, 
which we call the Fan \& Gijbels function (see \cite{fan95}). This function was already used in the volatility estimation context in \cite{gugu2020}. We generated data using the drift coefficient  $b(x) = -x$. For the noise level we took  $\eta_v=0.01$, which is substantial. Given the general difficulty of learning the volatility from noisy observations, one cannot expect to infer it on a very fine resolution (cf.~the remarks in Subsection~\ref{sec:prior}), and thus we opted for $N=40$ bins.
Experimentation showed that the results were rather robust w.r.t.\ the number of bins.

\begin{figure}
\begin{center}
\includegraphics[width=0.75\textwidth]{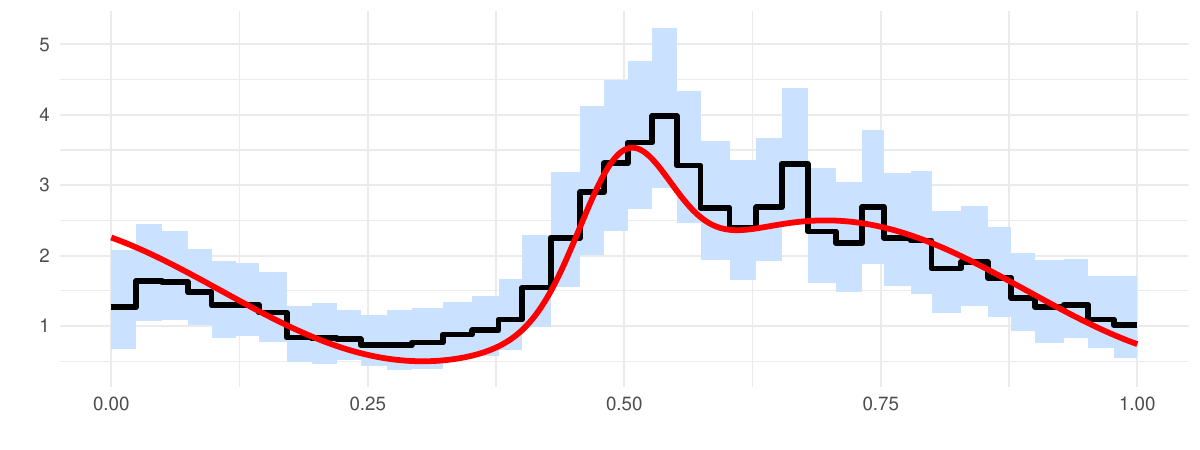}
\caption{Posterior and pointwise $95\%$ credible band for the example of Subsection~\ref{subsec:fan}. The true volatility function is plotted in red, the black step function gives the posterior mean, and the credible band is shaded in blue.}
\label{fig:fan}
\end{center}
\end{figure}

\begin{figure}
\begin{center}
\includegraphics[width=0.25\textwidth]{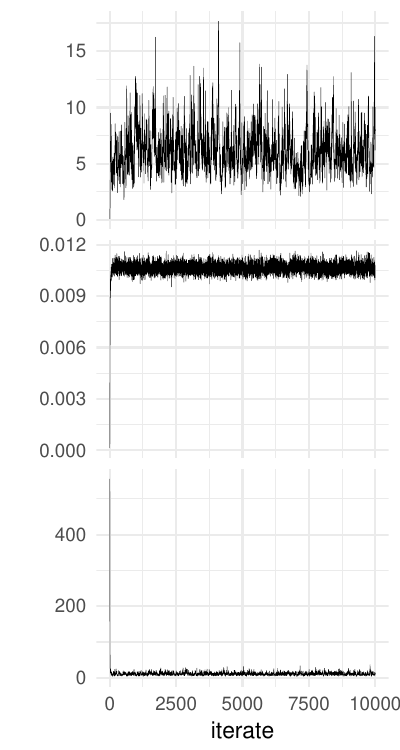}
\includegraphics[width=0.5\textwidth]{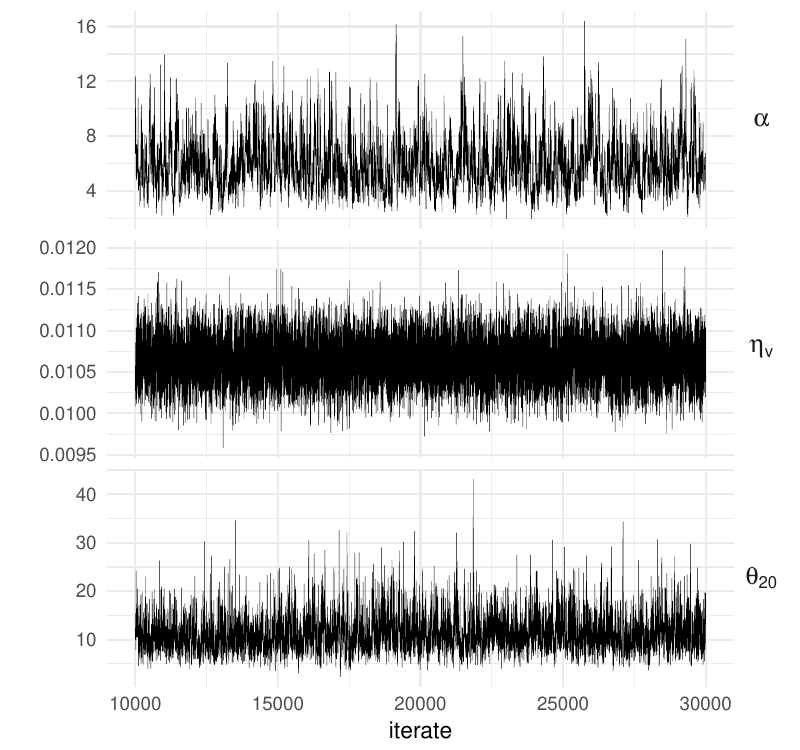}
\caption{Trace plots for the Fan \& Gijbels function of Subsection~\ref{subsec:fan}. Left: first $10.000$ samples. Right: subsequent $20.000$ samples. Top: $\alpha$, middle: $\eta_v$, bottom: $\th_{20}$.}
\label{fig:fan:trace}
\end{center}
\end{figure}

Inference results are reported in Figure~\ref{fig:fan}. It can be seen from the plot that we succeed in learning the volatility function. Note that the credible band does not cover the entire volatility function, but this had to be expected given that this is a marginal band. 
Quick mixing of the Markov chains can be visually assessed via the trace plots in Figure~\ref{fig:fan:trace}. 
The algorithm took about~11 minutes to complete.

\subsection{Heston model}\label{subsec:heston}
The Heston model (see~\cite{heston93}, or \cite[Section~10.3.3]{filipovic2009} and \cite[Chapter~19, Appendix~A]{brigomercurio2006}) is  a widely used stochastic volatility model. It stipulates that  the price process of a certain asset, denoted by $S$, evolves over time according to the SDE
\[
\dd S_t = \mu S_t \dd t + \sqrt{Z_t } S_t \dd W_t,
\]
where the process $Z$ follows the CIR or square root process (see \cite{cox85}),
\begin{equation}\label{eq:ZCIR}
\dd Z_t = \kappa ( \theta - Z_t ) \dd t + \sigma \sqrt{Z_t} \dd B_t.
\end{equation}
Here $W$ and $B$ are correlated Wiener processes with correlation $\rho$. Following a usual approach in quantitative finance, we work not with $S$ directly, but with its logarithm $X_t=\log S_t$. According to It\^{o}'s formula it obeys a diffusion equation with volatility $s(t)=\sqrt{Z_t}$,
\[
\dd X_t = (\mu-\frac{1}{2} Z_t)\dd t+\sqrt{Z_t}\dd W_t.
\]
We assume high-frequency observations on  the log-price process $X$ with additive noise $V_i \sim N(0,\eta_v)$. In this setup the volatility $s$ is a random function. We assume no further knowledge of the data generation mechanism. This setting is extremely challenging and puts our method to a serious test. To generate data, we used the parameter values
$\mu = 0.05,$
$\kappa = 7,$
$\theta = 0.04,$
$\sigma = 0.6,$
$\rho = -0.6$,
that mimick the ones obtained via fitting the Heston model to real data  (see Table 5.2 in \cite{vdp06}). Furthermore, the noise variance was taken equal to $\eta_v = 10^{-6}.$ The latter choice ensures a sufficiently large signal-to-noise ratio in the model \eqref{eq:ssm}, that can be quantified via the ratio $w_i / \eta_v$ of the \emph{intrinsic} noise level $w_i$ to the \emph{external} noise level $\eta_v$. Finally, the parameter choice $\mu=0.04$ corresponds to a $4\%$ log-return rate, which is a reasonable value.

\begin{figure}
\begin{center}
	\includegraphics[width=0.75\textwidth]{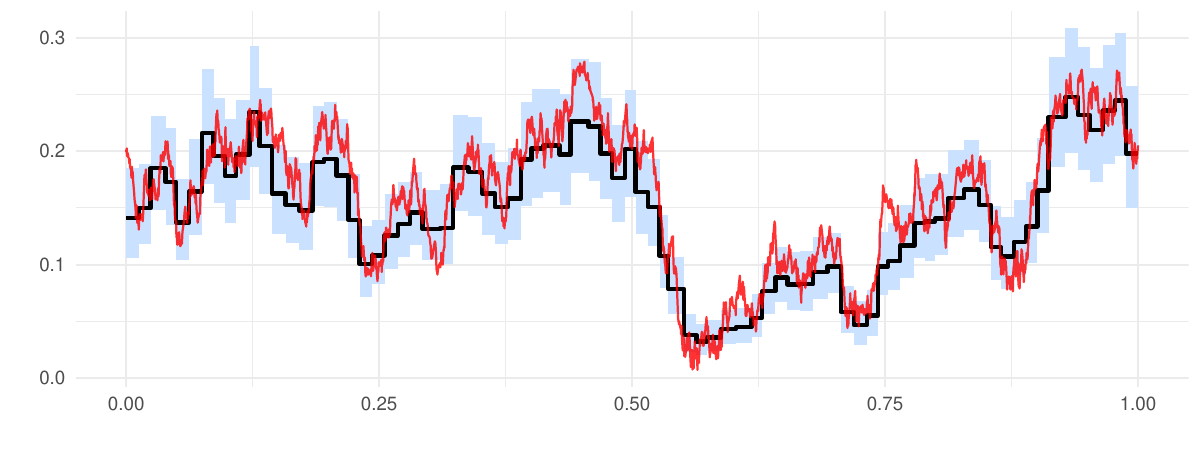}
	\caption{Posterior mean and pointwise $95\%$ credible band for the example of Subsection~\ref{subsec:heston}.  The true volatility function is plotted in red, the black step function gives the posterior mean, and the credible band is shaded in blue.}
\label{fig:heston}
\end{center}
\end{figure}

\begin{figure}
\begin{center}
	\includegraphics[width=0.25\textwidth]{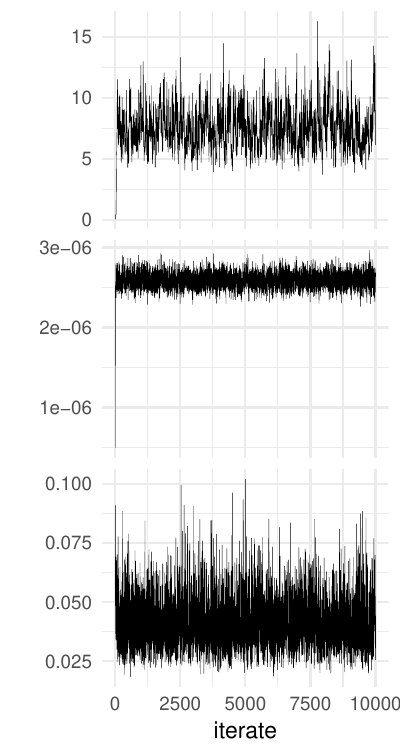}
		\includegraphics[width=0.5\textwidth]{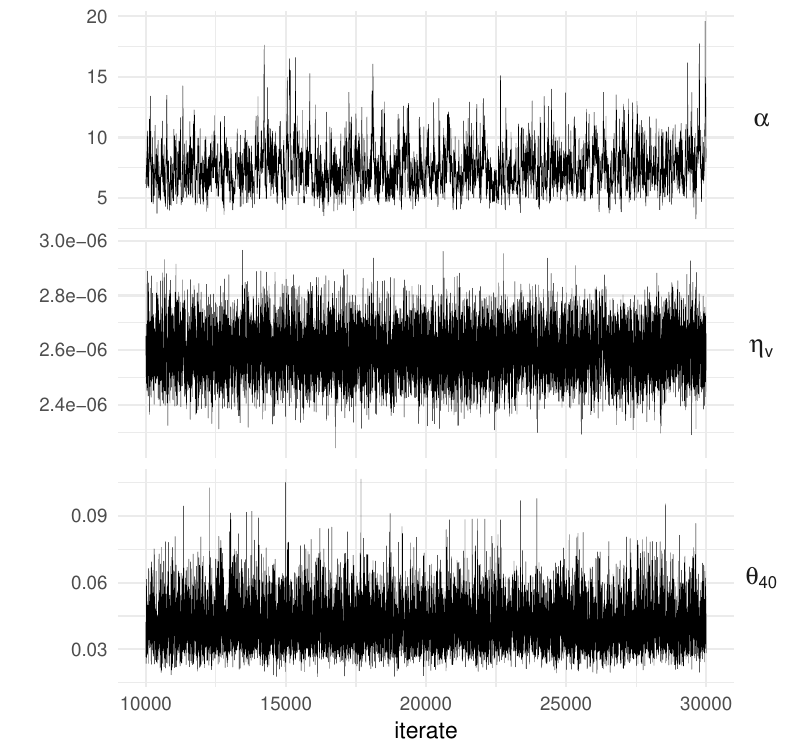}
	\caption{Trace plots for the Heston model of Subsection~\ref{subsec:heston}. Left: first $10.000$ samples. Right: subsequent $20.000$ samples. Top: $\alpha$, middle: $\eta_v$, bottom: $\th_{40}$.}
\label{fig:heston:trace}
\end{center}
\end{figure}

We report inference results with $N=80$ bins in Figure~\ref{fig:heston}. These are surprisingly accurate, given a general difficulty of the problem and the amount of assumptions that went into the learning procedure. 
 Note that the number of bins to get accurate results is higher than in the previous example. This is due to the fact that the volatility here is much rougher, H\"older continuous of order less than $\tfrac{1}{2}$. 
The Markov chains mix rapidly, as evidenced by the trace plots in Figure~\ref{fig:heston:trace}. 
The simulation run took about~12 minutes to complete. Finally, we note that the Heston model does not formally fall into the framework under which our Bayesian method was derived (deterministic volatility function $s$). We expect that a theoretical validation why our approach  also works  in this case requires the use of intricate technical arguments, which lie beyond the scope of the present, practically-oriented paper.

\begin{figure}
\begin{center}
	\includegraphics[width=0.75\textwidth]{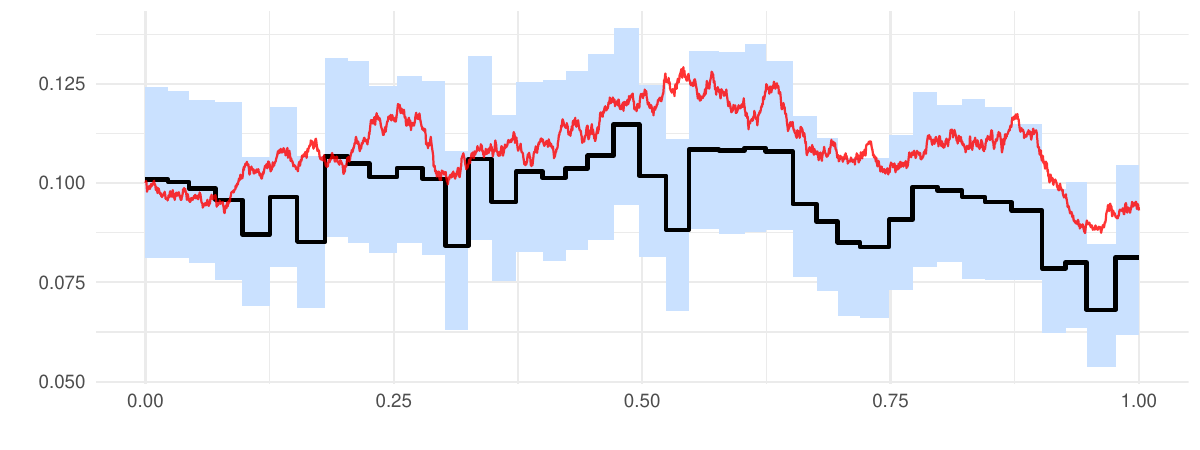}
	\caption{Posterior results for the Heston model where the CIR volatility  is replaced by the root of the continuous time limit of the  prior, $N=40$, starting value of the volatility is $0.1$. The colors have the same meaning as in Figure~\ref{fig:heston}.}
\label{fig:var_posterior_start40_4}
\end{center}
\end{figure}

\begin{figure}
\begin{center}
	\includegraphics[width=0.75\textwidth]{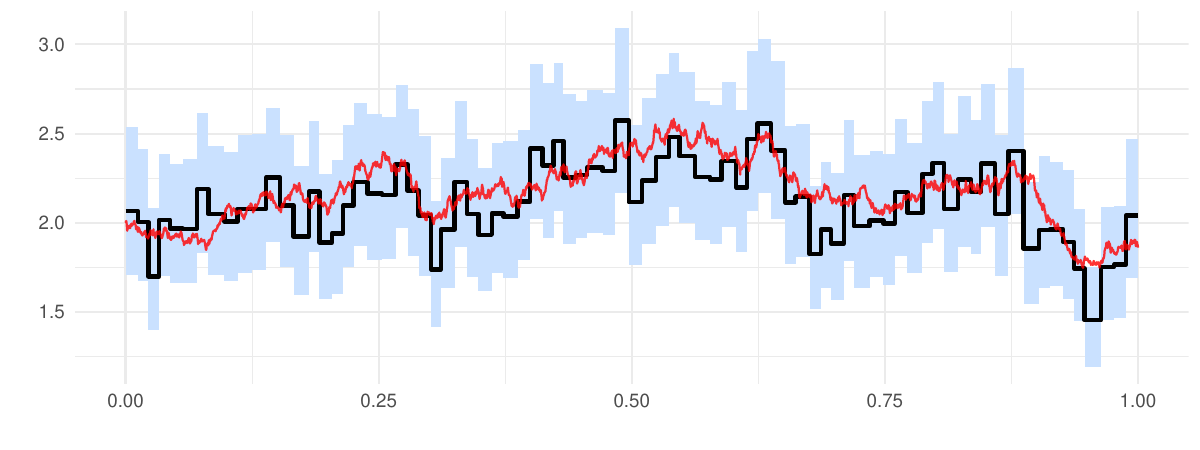}
	\caption{Posterior results for the Heston model where the CIR volatility  is replaced by the root of the continuous time limit of the  prior, $N=80$, starting value of the volatility is $2.0$.}
\label{fig:var_posterior_start80_4}
\end{center}
\end{figure}

We have also run some experiments for a variation of the above Heston model with the original CIR squared volatility according to \eqref{eq:ZCIR} replaced with a squared volatility equal to the limit process $\theta_1 Z$, where  $Z$ is as in \eqref{eq:limitZ}, equivalently as in \eqref{eq:sdez}, with $\sqrt{\frac{2}{\gamma}}=0.6$. The experiments showed the influence of the number of bins $N$ and the starting values, illustrated by Figures~\ref{fig:var_posterior_start40_4} and \ref{fig:var_posterior_start80_4}. One sees that the results with the lower number of bins $N=40$ combined with a low initial value are less satisfactory than with the higher $N=80$ and higher starting value.

\section{Exchange rate data}
\label{section:real}

Unlike daily stock quotes or exchange rate series that can easily be obtained via several online resources (e.g., Yahoo Finance), high frequency financial data are rarely accessible for free for academia. In this section, we apply our methodology to infer volatility of the high frequency foreign exchange rate data made available by Pepperstone Limited, the London based forex broker.\footnote{As of 2020, data are not available from the Pepperstone website any more, but can be obtained directly from the present authors.
%See \url{https://pepperstone.com/uk/client-resources/historical-tick-data} (Accessed on 25 April 2018).
The data are stored as csv files, that contain the dates and times of transactions and bid and ask prices. The data over 2019 are available for download (after a free registration) at \url{https://www.truefx.com/truefx-historical-downloads}.
} As we shall see below, the inferred volatility looks plausible, and while there is substantial uncertainty surrounding the inferential results left (as quantified by the marginal credible band), nontrivial conclusions can nevertheless be drawn.

\begin{figure}
\begin{center}
	\includegraphics[width=0.75\textwidth]{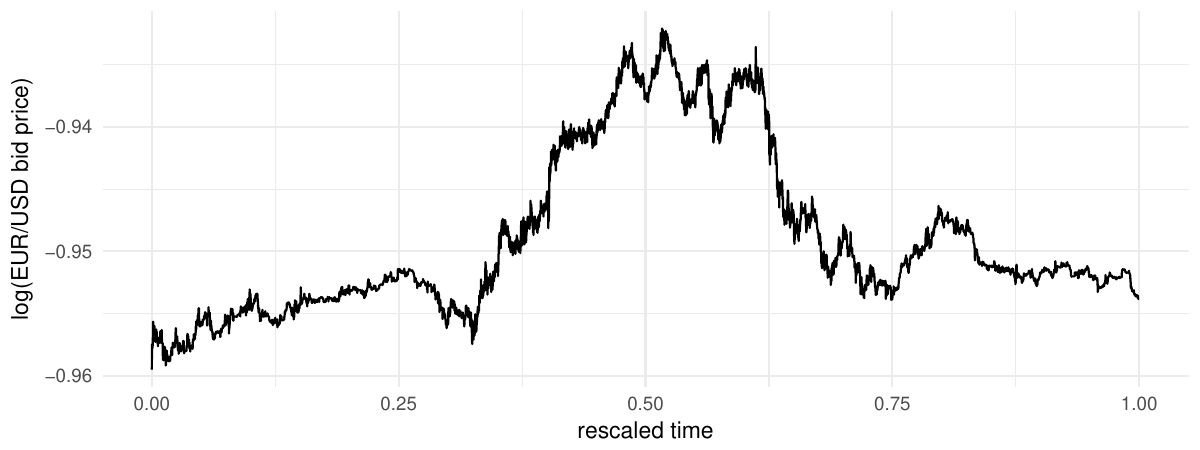}
	\includegraphics[width=0.75\textwidth]{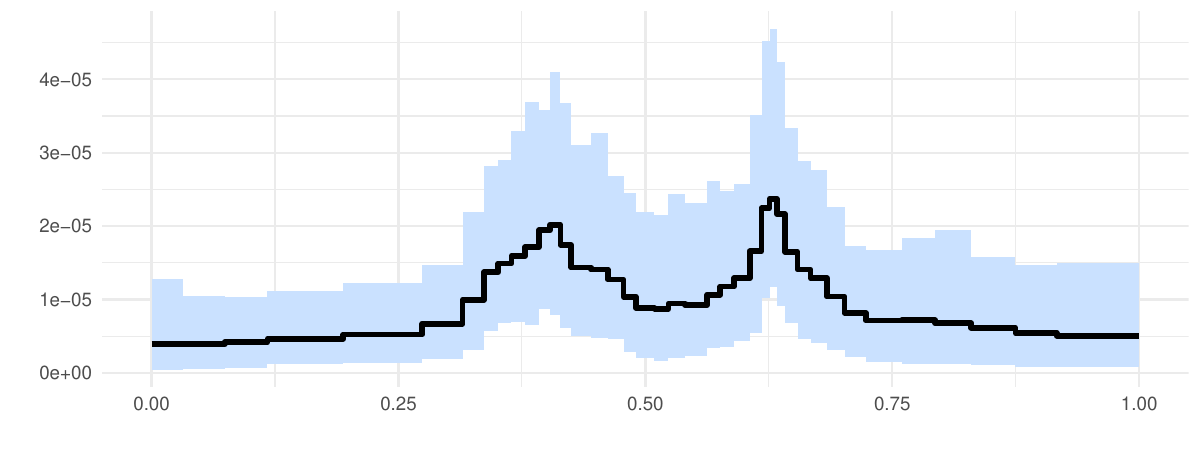}
	\caption{Top: Natural logarithm of the EUR/USD exchange rate data for 2 March 2015 analysed in Section~\ref{section:real}. Bottom: Posterior mean (black curve) and pointwise 95\% credible band (blue band) for the volatility function. The time axis is rescaled to $[0,1]$.}
\label{fig:eurus:data}
\end{center}
\end{figure}

Specifically, we use the EUR/USD tick data (bid prices) for 2 March 2015. As the independent additive measurement error model \eqref{eq:obs} becomes harder to justify for highly densely spaced data, we work with the subsampled data, retaining every $10$th observation. Applying various amounts of subsampling is a common strategy in this context; see, e.g., Section 5 in \cite{mancini2015spot} for a practical example. In our case subsampling results in a total of $n=13\, 025$ observations over one day, about~9 per minute. 
See \cite[Section~2.5]{mykland12} for further motivation and explanation of subsampling and \cite[Section~1.2]{zhang2005tale} where it is shown that subsampling (and averaging) is motivated by a substantial decrease in the bias of their estimator.
As in Subsection~\ref{subsec:heston}, we apply the log-transformation on the observed time series, and assume the additive measurement error model \eqref{eq:obs}. The data are displayed in Figure~\ref{fig:eurus:data}, top panel.

Financial time series often contain jumps, accurate detection of which is a delicate task. As this section serves illustrative purposes only, for simplicity we ignore jumps in our analysis. For high frequency data, volatility estimates are expected to recover quickly after infrequent jumps in the underlying process. This should in particular be the case for our learning procedure, given that we model the volatility as a piecewise constant function, which can be viewed as an estimate of the ``histogramised'' true volatility. Indeed, effects of infrequent jumps in the underlying process are likely to get smoothed out by averaging (we make no claim of robustness of our procedure with respect to possible presence of large jumps in the process $X$, or densely clustered small jumps). An alternative here would have been to use a heuristic jump removal technique, such as the ones discussed in \cite{sabel15}; next one could have applied our Bayesian procedure on the resulting ``cleaned'' data.

\begin{figure}
\begin{center}
	\includegraphics[width=0.75\textwidth]{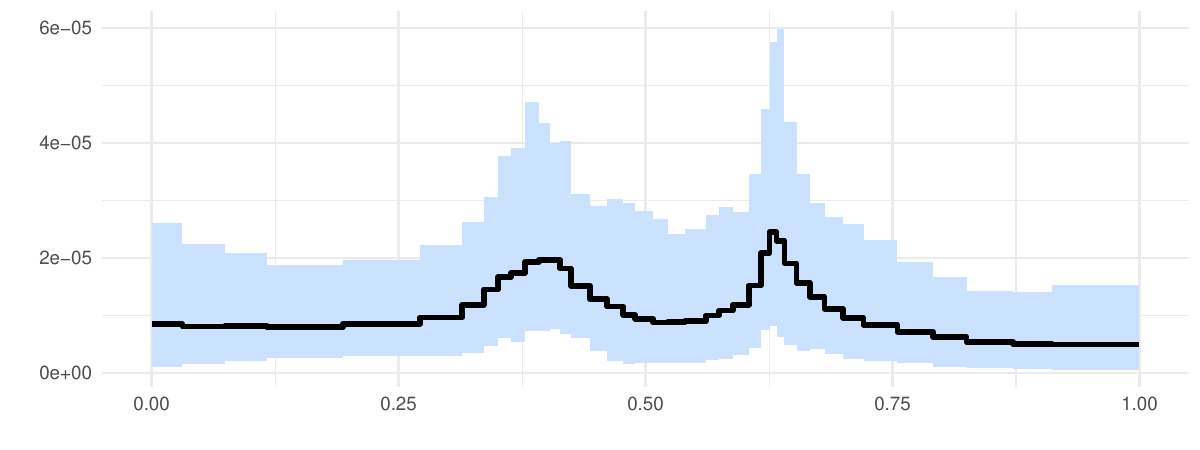}
	\caption{Posterior mean (black curve) and pointwise 95\% credible band (blue band) for the volatility of the further subsampled EUR/USD exchange rate data analysed in Section~\ref{section:real}. The time axis is rescaled to $[0,1]$.}
\label{fig:eurus2}
\end{center}
\end{figure}

\begin{figure}
\begin{center}
	\includegraphics[width=0.75\textwidth]{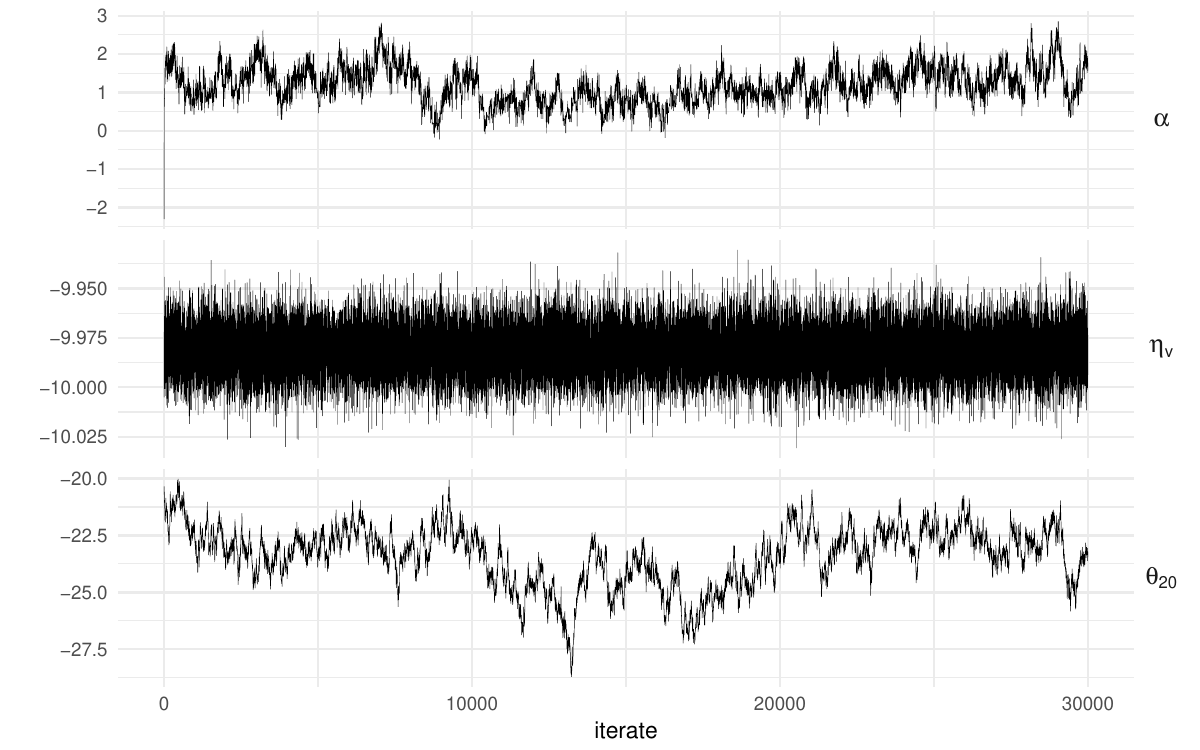}
	\caption{Log trace plots for the exchange rate data example of Section~\ref{section:real}.  Top: $\alpha$, middle: $\eta_v$, bottom: $\th_{20}$. }
\label{fig:eurus:trace}
\end{center}
\end{figure}

We report inference results in the bottom panel of Figure~\ref{fig:eurus:data}, where time is rescaled to the interval $[0,1]$, so that $t=0$ corresponds to the midnight and $t=0.5$ to  noon.  We used $N=96$ bins. As seen from the plot, there is considerable uncertainty in volatility estimates. Understandably, the volatility is lower during the night hours. It also displays two peaks corresponding to the early morning and late afternoon parts of the day. Finally, we give several trace plots of the generated Markov chains in Figure~\ref{fig:eurus:trace}. The algorithm took about~33 minutes to complete. In Figure~\ref{fig:eurus2} we give inference results obtained via further subsampling of the data, retaining $50\%$ of the observations. The posterior mean is quite similar to that in Figure~\ref{fig:eurus:data}, whereas the wider credible band reflects greater inferential uncertainty due to a smaller sample size. The figure provides a partial validation of the model we use.

\section{Discussion}
\label{section:conclusions}

In this paper we studied a practical nonparametric Bayesian approach to volatility learning under  microstructure noise. From the statistical theory point of view, the problem is much more difficult than volatility learning from noiseless observations. Hence, accurate inference on volatility under  microstructure noise requires large amounts of data. Fortunately, these are available in financial applications. On the other hand, design of a learning method that scales well with data becomes important.
Our specification of the prior and a deliberate, but asymptotically harmless misspecification of the drift by taking $b\equiv 0$ are clever choices that enable us to combine our earlier work in \cite{gugu18} with the FFBS algorithm for Gaussian linear state space models. This gives a conceptually simple and fast algorithm (Gibbs sampler) to obtain samples from the posterior, from which inferences can be drawn in a straightforward way.
A very interesting topic for future research within our approach is to explicitly account for the possible presence of jumps in financial time series.

\renewcommand{\thesection}{\Alph{section}}
\setcounter{section}{0}

\section{Details on update steps in the Gibbs sampler}
\label{appendixB}

\subsection{Drawing $x_{0:n}$}

We first describe how to draw the state vector $x_{0:n}$ conditional on all other parameters in the model and the data $y_{1:n}$. Note that for $u_i$ in \eqref{eq:ssm} we have by \eqref{ui} that $u_i \sim N(0,w_i)$, where
\begin{equation}
\label{wi}
\begin{split}
w_i &= \theta_k\Delta_i, \quad i \in [(k-1)m+1,km], \quad k=1,\ldots,N-1,\\
w_i &=  \theta_N \Delta_i, \quad i \in [(N-1)m+1,n].
\end{split}
\end{equation}
By equation (4.21) in \cite{petris09} (we omit dependence on $\theta_{1:N}, \eta_v$ in our notation, as they stay fixed in this step),
\[
p(x_{0:n} | y_{1:n})=\prod_{i=0}^n p( x_i | x_{i+1:n},  y_{1:n}),
\]
where the  factor with $i=n$ in the product on the righthand side is the filtering density $p( x_n | y_{1:n})$. This distribution  is in fact $N(\mu_n,C_n)$, with the mean $\mu_n$ and variance $C_n$ obtained from Kalman recursions
\[
\mu_i=\mu_{i-1}+K_i e_i,\quad C_i=K_i \eta_v, \quad i=1,\ldots,n.
\]
Here
\[
K_i =\frac{C_{i-1}+w_i }{C_{i-1} + w_i + \eta_v}, \quad i=1,\ldots,n,
\]
is the \emph{Kalman gain}. Furthermore,
$
e_i=y_i-\mu_{i-1}
$
is the one-step ahead \emph{prediction error}, also referred to as \emph{innovation}. See \cite{petris09}, Section 2.7.2. This constitutes the forward pass of the FFBS.

Next, in the backward pass, one draws backwards in time $\widetilde{x}_n \sim N(\mu_n,C_n)$ and  $\widetilde{x}_{n-1},\ldots \widetilde{x}_0$ from the densities $p( x_i | \widetilde{x}_{i+1},  y_{1:n})$ for $i=n-1,n-2,\ldots,0$. It holds that $p( x_i | \widetilde{x}_{i+1:n},  y_{1:n})=p( x_i | \widetilde{x}_{i+1},  y_{1:n})$, and the latter distribution is $N(h_i,H_i)$, with
\[
h_i=\mu_i+\frac{C_i}{C_i+w_{i+1}}(\widetilde{x}_{i+1}-\mu_i), \quad
H_i 
=\frac{C_i w_{i+1}}{C_i+w_{i+1}}.
\]
For every $i$, these expressions depend on a previously generated $\widetilde{x}_{i+1}$ and other known quantities only. The sequence $\widetilde{x}_0,\widetilde{x}_1,\ldots,\widetilde{x}_n$ is a sample from $p(x_{0:n}|y_{1:n})$. See Section 4.4.1 in \cite{petris09} for details on FFBS.

\subsection{Drawing $\eta_v$, $\theta_{1:N}$ and $\zeta_{2:N}$}\label{section:drawing}

Using the likelihood expression from Subsection~\ref{section:likelihood} and the fact that $\eta_v\sim\ig(\alpha_v,\beta_v)$, one sees that the full conditional distribution of $\eta_v$ is given by
\[
\eta_v | x_{1:n}, y_{1:n} \sim \ig\left( \alpha_v + \frac{n}{2}, \beta_v + \frac{1}{2} \sum_{i=1}^n (y_i-x_i)^2 \right).
\]
Similarly, using the likelihood expression from Subsection~\ref{section:likelihood} and the conditional distributions in \eqref{formula:prior}, one sees that the full conditional distributions for $\theta_{1:N}$ are
\begin{align*}
\theta_1 | \zeta_2,x_{1:n} &\sim \mathrm{IG}\left(\alpha_1 + \alpha +\frac{m_1}{2},\, \beta_1 +  \frac{\alpha}{ \zeta_{2}} + \frac{ Z_1}{2}\right),\\
\theta_k | \zeta_k,\zeta_{k+1},x_{1:n} &\sim \mathrm{IG}\left(2\alpha+\frac{m_k}{2},\, \frac{\alpha}{\zeta_k}+\frac{\alpha}{\zeta_{k+1}} + \frac{ Z_k}{2}\right), \quad k=2,\dots,N-1, \\
\theta_N | \zeta_N,x_{1:n} &\sim  \mathrm{IG}\left(\alpha+\frac{m_N}{2},\, \frac{\alpha}{ \zeta_N} + \frac{ Z_N}{2}\right).
\end{align*}
The full conditional distributions for $\zeta_{2:N}$ are
\[
\zeta_k | \theta_k,\theta_{k-1} \sim \ig\left(2\alpha,\frac{\alpha}{\theta_{k-1}}+\frac{\alpha}{ \theta_k}\right), \quad k=2,\ldots,N.
\]

\subsection{Drawing $\alpha$}

The unnormalised full conditional density of $\alpha$ is
\[
q(\alpha) = \pi(\alpha)\left(\frac{\alpha^{\alpha}}{\Gamma(\alpha)}\right)^{2(N-1)} \exp\left(-\alpha\sum_{k=2}^N \frac{1}{\zeta_k}\left( \frac{1}{\theta_{k-1}} + \frac{1}{\theta_k}\right)\right)  \prod_{k=2}^N \left(\theta_{k-1} \theta_k \zeta_k^2 \right)^{-\alpha}.
\]
The corresponding normalised density is nonstandard, and the Metropolis-within-Gibbs step (see, e.g., \cite{tierney94}) is used to update $\alpha$. The specific details are exactly the same as in \cite{gugu18}.

\subsection{Gibbs sampler}
\label{appendixC}

Settings for the Gibbs sampler in Section~\ref{section:simulations} are as follows: we used a vague specification $\alpha_1,\beta_1\rightarrow 0$, and also assumed that $\log\alpha\sim N(1,0.25)$ and $\eta_v \sim \ig(0.3,0.3)$ in Subsection~\ref{subsec:fan}. For the Heston model in Subsection~\ref{subsec:heston} we used the specification $\eta_v \sim \ig(0.001,0.001)$. 
Furthermore, we set $x_0 \sim N(0,25)$. The Metropolis-within-Gibbs step to update the hyperparameter $\alpha$ was performed via an independent Gaussian random walk proposal (with a correction as in \cite{wilkinson12}) with scaling to ensure the acceptance rate of about~$30-50\%$. The Gibbs sampler was run for $30 \, 000$ iterations, with the first third of the samples dropped as  burn-in.

%%%%%%%%%%%%%%%%%%%

%--------------------
\section*{Declarations}

\subsection*{Funding}
The research leading to the results in this paper has received funding from the European Research Council under ERC Grant Agreement 320637.

\subsection*{Competing interests}
The authors have no competing interests to declare that are relevant to the content of this article.

\subsection*{Code availability}
The computer code to reproduce numerical examples in this article is available at \cite{microstructure22}.

%\begin{acknowledgement}
%The research leading to the results in this paper has received funding from the European Research Council under ERC Grant Agreement 320637.
%\end{acknowledgement}

\bibliographystyle{spmpsci}
 \bibliography{bibliography}

\end{document}